\documentclass{aastex631}
\usepackage{xcolor}

\newcommand{\derfrac}[2]{\frac{\text{d} #1}{\text{d} #2}}

\newif\ifrev
\revfalse

\newcommand{\rev}[1]{\ifrev\leavevmode{\bf#1}\else #1\fi}

\begin{document}

\title{Linearity of Structure Kernels in Main-sequence and Subgiant Solar-like Oscillators}

\correspondingauthor{Lynn Buchele}
\email{lynn.buchele@h-its.org} 

\author[0000-0003-1666-4787]{Lynn Buchele} 
\affiliation{Heidelberg Institute for Theoretical Studies, Schloss-Wolfsbrunnenweg 35, 69118 Heidelberg, Germany}
\affiliation{Center for Astronomy (ZAH/LSW), Heidelberg University, Königstuhl 12, 69117 Heidelberg, Germany}

\author[0000-0003-4456-4863]{Earl P.~Bellinger}
\affiliation{Department of Astronomy, Yale University, PO Box 208181, New Haven, CT 06520-8101, USA}

\author[0000-0002-1463-726X]{Saskia Hekker}
\affiliation{Heidelberg Institute for Theoretical Studies, Schloss-Wolfsbrunnenweg 35, 69118 Heidelberg, Germany}
\affiliation{Center for Astronomy (ZAH/LSW), Heidelberg University, Königstuhl 12, 69117 Heidelberg, Germany}

\author[0000-0002-6163-3472]{Sarbani Basu}
\affiliation{Department of Astronomy, Yale University, PO Box 208181, New Haven, CT 06520-8101, USA}

\begin{abstract}
Seismic structure inversions have been used to study the solar interior for decades. With the high-precision frequencies obtained using data from the \emph{Kepler} mission, it has now become possible to study other solar-like oscillators using structure inversions, including both main-sequence and subgiant stars. Subgiant stars are particularly interesting because they exhibit modes of mixed acoustic-buoyancy nature, which provide the opportunity to probe the deeper region of stellar cores. This work examines whether the structure inversion techniques developed for the pure acoustic modes of the Sun and other main-sequence stars are still valid for mixed modes observed in subgiant stars. We construct two grids of models: one of main-sequence stars and one of early subgiant stars. Using these grids, we examine two different parts of the inversion procedure. First, we examine what we call the ``kernel errors", which measure how well the mode sensitivity functions can recover known frequency differences between two models. Second, we test how these kernel errors affect the ability of an inversion to infer known structure differences. On the main sequence, we find that reliable structure inversion results can be obtained across the entire \rev{range of masses and large frequency separations we consider}. On the subgiant branch, however, the rapid evolution of mixed modes leads to large kernel errors and hence difficulty recovering known structure differences. Our tests show that using mixed modes to infer the structure of subgiant stars reliably will require improvements to current fitting approaches and modifications to the structure inversion techniques.

\end{abstract}


\section{Introduction} \label{sec:intro}
The field of asteroseismology, i.e., the study of stars through their global oscillations, has provided unprecedented insight into stellar interiors \citep[e.g.,][]{2021RvMP...93a5001A, 2024arXiv241001715B}.  One of the most powerful techniques in asteroseismology is that of structure inversions. This technique utilizes the inherent sensitivity of each oscillation mode to infer information about structure differences between a star and its  model (typically a best-fit model found through a grid-based or optimization procedure). These inferred differences are then used to assess how well the internal structure of a star is reproduced by a given stellar model as well as to explore potential improvements to stellar modeling.  The roots of this approach lie in the field of geology, which use oscillations excited by earthquakes to infer the interior structure of the Earth \citep{1968GeoJ...16..169B, 1970RSPTA.266..123B}. The techniques were then adapted by helioseismologists to study the internal structure of the Sun \citep[e.g.][]{1991sia..book..519G, 2002RvMP...74.1073C, 2016LRSP...13....2B}. 

In anticipation of high-quality asteroseismic data, helioseismologists began testing what changes would be necessary to apply structure inversions to stars other than the Sun \citep{1993ASPC...40..541G, 2002ESASP.485..249B, 2003Ap&SS.284..153B}. Much of this work focused on adapting existing helioseismic techniques to the higher uncertainties and reduced number of modes that can be observed in other stars. 
Despite initial observations of solar-like oscillations in other stars from ground-based campaigns \citep{1991ApJ...368..599B,1995AJ....109.1313K, 2001ApJ...549L.105B, 2007CoAst.150..106B} and early space-based missions such as CoRoT \citep{2009A&A...506..411A, 2009IAUS..253...71B},  the first asteroseismic targets to be probed with localized structure inversions \citep{2017ApJ...851...80B} were targets of the \emph{Kepler} Space Telescope \citep{2010Sci...327..977B, 2010ApJ...713L..79K}. This is because structure inversions require a high number of individual modes to be precisely identified. 

In this work, we focus specifically on linear localized structure inversions that rely on functions known as stellar structure kernels \citep[e.g.][]{2017ApJ...851...80B, 2019ApJ...885..143B, 2020IAUS..354..107K, 2021ApJ...915..100B, 2022A&A...661A.143B, 2023A&A...675A..17V,  2024ApJ...961..198B, 2024arXiv241205094B}. However, similar inverse techniques are used to infer a broader range of internal properties of an observed star. These internal properties can include global indicators \citep[e.g.][]{2012A&A...539A..63R, 2015A&A...574A..42B, 2015A&A...583A..62B, 2018A&A...609A..95B}, rotational profiles \citep[e.g.][]{ 2014MNRAS.444..102K, 2014A&A...564A..27D, 2015MNRAS.452.2654B,2017A&A...602A..62T,2020A&A...639A..98A, 2022A&A...668A..98A, 2025A&A...693A.274A}, and more recently, the strength of internal magnetic fields \citep{2022Natur.610...43L, 2023A&A...680A..26L}.

Using data from \emph{Kepler}, initial targets of asteroseismic inversions included
the solar analogs 16~Cyg~A~and~B \citep{2017ApJ...851...80B, 2022A&A...661A.143B}, a star with a small convective core \citep{2019ApJ...885..143B}, and several more evolved stars exhibiting mixed modes \citep{2020IAUS..354..107K, 2021ApJ...915..100B}. Recent works \citep{2024ApJ...961..198B,2024arXiv241205094B} have examined all the stars in \citet{2016MNRAS.456.2183D} and \citet{2017ApJ...835..172L} and found that 55 of these stars could be studied with localized structure inversions. 

While this is a large improvement over the previous sample of one star, the Sun, this still relatively small sample size makes it difficult to draw conclusions on potential improvements to stellar modeling \citep{2020IAUS..354..107K, 2024ApJ...961..198B, 2024arXiv241001715B}. Unfortunately, the requirement for the highest quality observations means that the number of stars that can be studied with these techniques will remain low until the next generation of asteroseismic missions, such as the PLATO mission \citep{2014ExA....38..249R, 2024arXiv240605447R}. Another limitation of the existing set of stars that have been studied with structure inversions is that they are restricted to a narrow range of masses and are still primarily on the main sequence. Thus, in order to truly probe stellar evolution theory, it is important to expand the types of stars that can be studied in this manner. However, care must be taken to ensure that the assumptions made in the process of such inversions hold true for stars with different structure and oscillation properties. 

Here we are specifically interested in the reliability of structure inversions for subgiant branch stars. As stars run out of hydrogen and leave the main sequence, they enter a short-lived evolutionary stage known as the subgiant branch. During this phase of evolution, the core of the star contracts. This raises the buoyancy frequency in the core. At the same time, the envelope of the star expands, decreasing the frequencies of modes excited by solar-like oscillations. These two effects result in the excitation of modes with two oscillation cavities: an outer cavity where modes propagate as pressure modes (p-modes) and an inner cavity where modes propagate as buoyancy modes (g-modes). In subgiant and red giant branch stars, the region between these two cavities is small and hence, some non-radial modes are able to propagate in both cavities. These modes have a mixed acoustic-buoyancy character and hence are known as mixed modes \citep{1975PASJ...27..237O, 2011A&A...535A..91D}. 

The degree to which any given mixed mode is predominantly acoustic or buoyant in nature changes as the star evolves. Mixed modes undergo a series of avoided crossings where the mode character is exchanged with a neighboring mode \citep{1977A&A....58...41A}. Thus, over the course of the subgiant branch any given mode will be alternatively p-dominated (mostly acoustic mode character) and g-dominated (mostly buoyancy mode character). All mixed modes are sensitive to the structure of the core in a way that differs from the pure acoustic modes observed in main-sequence solar-like oscillators. This sensitivity make subgiant stars appealing targets for structure inversions as mixed modes mean that structure inversions can probe deep in the core. 

This increased sensitivity has been used in a few works which apply structure inversions to stars which exhibit mixed modes in their oscillation spectra. First, \citet{2020IAUS..354..107K} presented results for two stars which they claimed exhibited mixed modes, although this mode identification has been questioned by \citet{2024arXiv241205094B}. Then, \citet{2021ApJ...915..100B} presented structure inversion results for HR~7322, one of the best-characterized subgiant stars observed by the \emph{Kepler} mission \citep{2019MNRAS.489..928S, 2020MNRAS.499.2445H}. However, neither of these works examined in detail the reliability of structure inversions for stars exhibiting mixed modes. Due to the rapid evolution of mixed modes both in frequency and in character, it is reasonable to question the validity of the assumptions made in structure inversions in the subgiant regime. This work seeks to explore these questions by comparing the reliability of structure inversions of main-sequence solar-like oscillators with solar-like oscillators on the subgiant branch. We begin in Section~\ref{sect:kern_inv} with a more detailed review of the process and assumptions of a structure inversion. Then in Section~\ref{sect:MS} we review the reliability of inversions while on the main sequence, before stars begin to exhibit mixed modes. Section~\ref{sect:SGB} extends this analysis to the subgiant branch, with a particular focus on mixed dipole modes. Finally, we summarize the work and present our conclusions in Section~\ref{sect:conc}.

\section{Kernels and Inversions} 
\label{sect:kern_inv} 
We begin here with a review of the process and assumptions of a structure inversion before defining the terms and tests used in the following sections. Structure inversions utilize the inherent sensitivity of oscillation mode frequencies to the underlying structure of the star. This sensitivity can be expressed in functions calculated from the structure variables of a stellar model called mode kernels. Mathematically, this is expressed as \citep{1990MNRAS.244..542D}
\begin{equation}
    \label{equ:mode_kern}
    \frac{\delta \nu_{i}}{\nu_{i}} = \int K_{i}^{(f_1, f_2)} \frac{\delta f_1}{f_1} \,\textrm{d}r + \int K_{i}^{(f_2,f_1)} \frac{\delta f_2}{f_2} \,\textrm{d}r,
\end{equation}
where the subscript \(i\) indicates a mode with a specific combination of radial order (\(n\)) and spherical degree (\(\ell\)), \(\nu_{i}\) is the corresponding mode frequency, \(K_{i}\) are the mode kernels, and \(f_1, f_2\) indicate the stellar structure variables being considered. The relative difference of a given quantity \(x\), which can represent either a mode frequency or structure variable, is \(\delta x/x \equiv (x_A - x_B)/x_B\), where typically \(A\) denotes properties of the star and \(B\) denotes properties of the model. In this work, however, we consider only comparisons between two models.

The mode kernels, \(K_{i}^{(f_1, f_2)}, K_{i}^{(f_2,f_1)}\) are derived through a perturbation of the variational formulation of the adiabatic eigenvalue problem \citep[e.g.,][]{1958HDP....51..353L}. Derivations of these kernels typically begin by deriving \(K_i\) in terms of the squared sound speed (\(f_1 = c^2\)) and density (\(f_2 = \rho\)) \citep[e.g.,][]{1991sia..book..519G, 1999JCoAM.109....1K}. Using the (\(c^2,\rho\)) kernels as a starting point, kernels of other variable pairs may be derived including: density and the first adiabatic exponent \(\Gamma_1\) \citep[e.g.,][]{1991sia..book..519G, 1993ASPC...40..541G}, squared isothermal sound speed \(u = c^2/\Gamma_1\) and helium mass fraction \(Y\) \citep[e.g.,][]{1997A&A...322L...5B, 1999JCoAM.109....1K, 2015A&A...583A..62B, 2017A&A...598A..21B},  and the convective stability parameter \(A\) and \(\Gamma_1\) \citep[e.g.,][]{10.1093/mnras/280.4.1244, 1999JCoAM.109....1K, 2017A&A...598A..21B}. For detailed discussions of how kernels are changed from one set of structure variables to another, see \citet{2011LNP...832....3K, 2002ESASP.485...95T, 2017A&A...598A..21B}. In this work, we focus specifically on the  (\(u,Y\)) variable pair as it is the most common pair used in asteroseismic inversions. For a more detailed comparison of the linearity of various pairs of structure variables we refer the reader to \citet{2017A&A...598A..21B}. 

There is another complication that arises when studying stars other than the Sun: the lack of precise values of the stellar mass, radius, and age. As the oscillation frequencies scale with the mean density of the star, a mismatch in mean density can introduce an error into the inversion results \citep{2003Ap&SS.284..153B}. To avoid this, it is common to perform asteroseismic inversions with respect to dimensionless variables, for example \rev{\(\hat{u} = u R /GM\) where \(R\) and \(M\) are the stellar radius and mass, respectively, and $G$ is the gravitational constant}. This requires the use of dimensionless frequencies:
\begin{equation}
    \label{equ:dimless_freq} 
    \hat{\nu} = \sqrt{\frac{R^3}{GM}} \, \nu
\end{equation}
where \(G\) is the gravitational constant. This scaling also means that the structure differences are computed at constant fractional radius $x = r/R$ rather than constant physical radius. For an observed star, where \(M\) and \(R\) are uncertain, the dimensionless frequency differences \(\delta \hat{\nu}/\hat{\nu}\) cannot be determined exactly. There are several different approaches to resolve this potential source of error, at various steps in the inversion process including in the inversion \rev{\citep{2020IAUS..354..107K, 2024MNRAS.527.1283T}}, while obtaining a best fit model \citep{2022A&A...661A.143B}, or during the computation of the frequency differences \citep{R98, 2003Ap&SS.284..153B,2021ApJ...915..100B,2024ApJ...961..198B}. Since the tests in this work are done by comparing two models, we opt to directly calculate \(\delta \hat{\nu} / \hat{\nu}\) using the known values of \(M\) and \(R\). 

While Equation~\ref{equ:mode_kern} describes the sensitivity of a single mode, one equation alone is insufficient to infer anything about the differences in structure because each oscillation mode is sensitive to many locations within the star. Therefore, \rev{one way forward is to infer localized differences using a linear combination of modes} constructed such that the resulting ``averaging" kernel has sensitivity only around a chosen target radius, \(r_0\). This is the method of optimally localized averages (OLA) \citep{1968GeoJ...16..169B, 1970RSPTA.266..123B}. Taking a linear combination of different modes turns Equation~\ref{equ:mode_kern} into: 
\begin{equation} 
\label{equ:kern_sum} 
    \sum_{i \in \mathcal{M}} c_{i} \frac{\delta \hat{\nu}_{i}}{\hat{\nu}_{i}} = \int \mathcal{K}(r_0) \frac{\delta \hat{u}}{\hat{u}} \, \textrm{d}x + \int \mathcal{C}(r_0) \delta Y \, \textrm{d}x.  
\end{equation} 
In this equation, \(\mathcal{M}\) represents the set of observed modes, \(c_{i}\) are the coefficients of the linear combination, called inversion coefficients, \(\mathcal{K}(r_0) = \sum_{i \in \mathcal{M}} c_{i} K_{i}^{(\hat{u},Y)}\) is the averaging kernel for a given target radius, \(\mathcal{C}(r_0) = \sum_{i \in \mathcal{M}} c_{i} K_{i}^{(Y,\hat{u})}\) is the cross-term kernel. \(\mathcal{K}\) is called an averaging kernel because if it is normalized to 1 and \(\mathcal{C}\) is small, then Equation~\ref{equ:kern_sum} becomes 
\begin{equation}
    \label{equ:avg_kern}    
    \sum_{i \in \mathcal{M}} c_{i} \frac{\delta \hat{\nu}_{i}}{\hat{\nu}_{i}} = \int \mathcal{K}(r_0) \frac{\delta \hat{u}}{\hat{u}} \, \textrm{d}x \approx \left<\frac{\delta \hat{u}}{\hat{u}} \right>_{r_0}, 
\end{equation}
and the linear combination of frequency differences can be interpreted as the localized average of the structure differences, where \(\mathcal{K}(r_0)\) is the weighting function. \rev{Assuming that the observational uncertainties are statistically independent of each other,} the uncertainty of an inversion result $\sigma_{\text{inv}}$ is given by the propagation of the uncertainties of the observed mode frequencies $\sigma_i$,
\begin{equation} 
    \label{equ:inv_uncertain} 
    \sigma_{\text{inv}} = \sum_{i\in \mathcal{M}} c^2_i \sigma_i^2. 
\end{equation} 

There are two common approaches to determine the coefficients which construct the localized averaging kernel: multiplicative optimally localized averages \citep[MOLA,][]{1968GeoJ...16..169B, 1970RSPTA.266..123B} and subtractive optimally localized averages \citep[SOLA,][]{1992A&A...262L..33P, 1994A&A...281..231P}. Here we provide a brief summary of each method. For a more detailed discussion of the implementation of each method, we refer readers to Chapter 10 of \citet{2017asda.book.....B}. 

MOLA constructs an averaging kernel by using a penalty function, which increases for points farther from the target radius, to suppress amplitude away from the target radius. This process requires the choice of two trade-off parameters: the cross-term suppression parameter and the error suppression parameter. These parameters are chosen to balance the localization of the averaging kernel, amplitude of the cross-term kernel, and amplification of the uncertainties.  SOLA begins by defining a target kernel, typically a modified Gaussian centered on the target radius, and constructs an averaging kernel that resembles the target kernel. In addition to the two trade-off parameters used in MOLA, SOLA also requires the choice of a parameter which determines the width of the target kernel. For our tests below, we choose by hand values of the trade-off parameters which result in nicely localized averaging kernels. For more details about choosing inversion parameters, see \citet{1999MNRAS.309...35R}.

\rev{In this work we use MOLA for our tests of models on the main-sequence, as was done in \citet{2024ApJ...961..198B} and  \citet{2024arXiv241205094B}. In our subgiant tests we use SOLA, as was the case in \citet{2021ApJ...915..100B}.} Regardless of the OLA method chosen, the procedure of the inversion remains the same. It begins by finding a reference model, a step typically referred to as forward modeling. This is typically the best-fit model resulting from a grid-based and/or optimization procedure. The structure of this reference model is then used to calculate the mode kernels that are then combined into an averaging kernel. Finally, the inversion coefficients are combined with the frequency differences between the observed star and the model to calculate the inferred structure difference. 

The key assumption of a structure inversion is that the mode kernels, calculated using a reference model, accurately represent the sensitivity of the observed modes \rev{to first-order}. This work focuses on the errors introduced by this assumption. Although we call these errors ``kernel errors'', it should be noted that these are not errors in the calculation of the mode kernels, rather they are errors that result from using a mode kernel that does not accurately represent the sensitivity of an observed mode to the underlying structure. Using Equation~\ref{equ:mode_kern}, we define the kernel error of a mode as
\begin{equation} 
    \label{equ:kern_err}
    {\text{Kernel Error}_i} \equiv  \frac{\delta \hat{\nu}_i}{\hat{\nu}_i} - \left[ \int K_i^{(\hat{u},Y)} \frac{\delta \hat{u}}{\hat{u}} \,\textrm{d}x + \int K_i^{(Y,\hat{u})} \delta Y \,\textrm{d}x\right]. 
\end{equation} 

Clearly, such a quantity cannot be computed for differences between an observed star and a model, and so our tests rely on comparisons between different stellar models. In each evolutionary stage we examine, we keep one model the same for all tests. We refer to this model as the reference model and denote any properties of it with the subscript `ref'. The second model we use we refer to as a test model, denoted with the subscript `test'. When computing kernel errors using Equation \ref{equ:kern_err} written in terms of our chosen variables $(\hat{u},Y)$, the mode kernels are those calculated using the reference model and the relative differences are defined as 
\begin{equation}
    \label{equ:rel_difs} 
    \frac{\delta \hat{\nu}}{\hat{\nu}} = \frac{\hat{\nu}_{\rm{test}} - \hat{\nu}_{\rm{ref}}}{\hat{\nu}_{\rm{ref}}} \quad {\rm{and}} \quad\frac{\delta \hat{u}}{\hat{u}} = \frac{\hat{u}_{\rm{test}} - \hat{u}_{\rm{ref}}}{\hat{u}_{\rm{ref}}} \quad {\rm{and}} \quad \delta Y = Y_{\rm{test}} - Y_{\rm{ref}}.
\end{equation}

In addition to examining how the kernel errors of different modes change across a grid of stellar models, we also explore how these kernel errors propagate through a full structure inversion. For this, we treat the test model as an observed star and use the reference model to perform structure inversions at several target radii. The result of these inversions can then be compared to the known structure differences between the two models. In this process, it can be helpful to also compare the inversion results to the true localized average difference, 
\begin{equation}
    \label{equ:lad} 
    \text{Localized Average Difference} \equiv \int \mathcal{K}(r_0) \frac{\delta \hat{u}}{\hat{u}} \, \textrm{d}x.
\end{equation}
This allows us to separate errors due to a poor averaging kernel or a large cross-term from those due to the underlying kernel errors. As structure inversions depend on the set of modes observed and their uncertainties, for each reference model we adopt a set of modes from an observed star with similar properties.  

\begin{figure} 
    \epsscale{0.5}
    \plotone{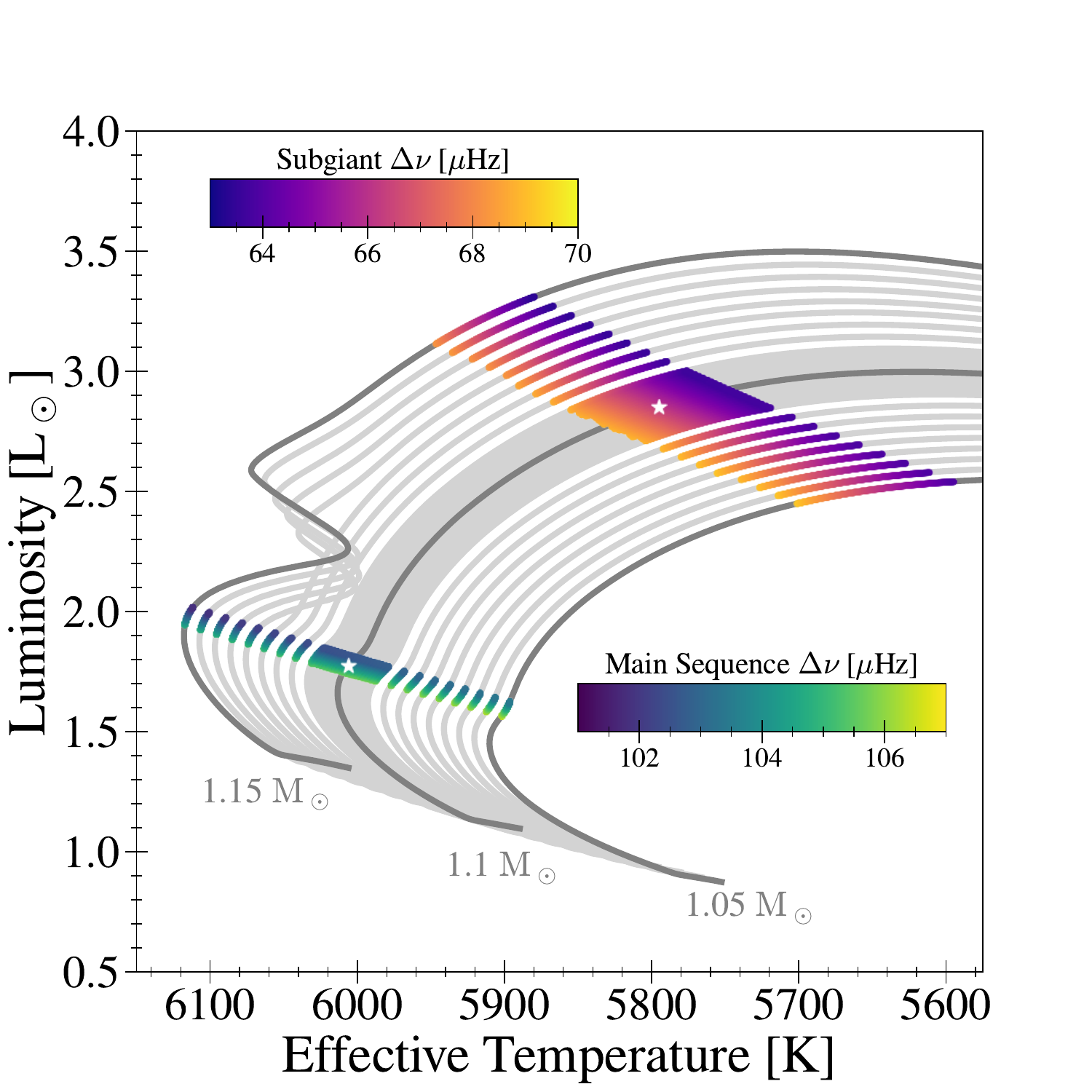} 
    \caption{Hertzsprung–Russell diagram showing the evolutionary tracks used in this work. For reference, we provide the masses of three tracks marked with darker gray lines. The colored points indicate the values of $\Delta \nu$ for models used to calculate the kernel errors, with each color map corresponding to a different evolutionary stage. The reference model used for each evolutionary stage is indicated with a white star.} 
    \label{fig:HRD}
\end{figure}

\rev{We applied these tests to two grids of models constructed using the MESA stellar evolution code \citep{2011ApJS..192....3P, 2013ApJS..208....4P, 2015ApJS..220...15P, 2019ApJS..243...10P, 2023ApJS..265...15J}. The only difference between each grid was the evolutionary stage under consideration. The masses we consider range from 1.05 to 1.15~M$_\odot$, which covers the $5\sigma$ range of the mass of the subgiant star $\mu$Her reported in \citet{2017ApJ...836..142G}. We varied only the mass and age of the models, keeping everything else (composition, mixing length, other input physics) constant. The initial composition is composition of $Y_{\rm{init}} = 0.28, Z_{\rm{init}} = 0.02$, where metal abundances have been scaled to the \citet{1998SSRv...85..161G} solar abundances.  In Figure~\ref{fig:HRD}, we show the evolutionary track used in our analysis and indicate the specific models used to analyze both the main sequence and subgiant evolutionary phases.}

\section{Main-Sequence Stars} 
\label{sect:MS} 
Before examining the behavior of structure kernels on the subgiant branch, we first explore the kernel behavior for a star on the main sequence. \rev{Using the grid parameters described above, we saved models every $5\times10^5$ yr along each evolutionary track},  while the frequency of maximum oscillation power ($\nu_{\rm{max}}$, calculated using scaling relations) was between 2520 and 2250~$\mu$Hz. We then used the GYRE oscillation code \citep{Townsend2013} to calculate the adiabatic eigenfrequencies and eigenfunctions. These were in turn then used to calculate the (\(\hat{u}, Y\)) structure kernels. 

\subsection{Kernel Errors} 
\begin{figure} 
    \plotone{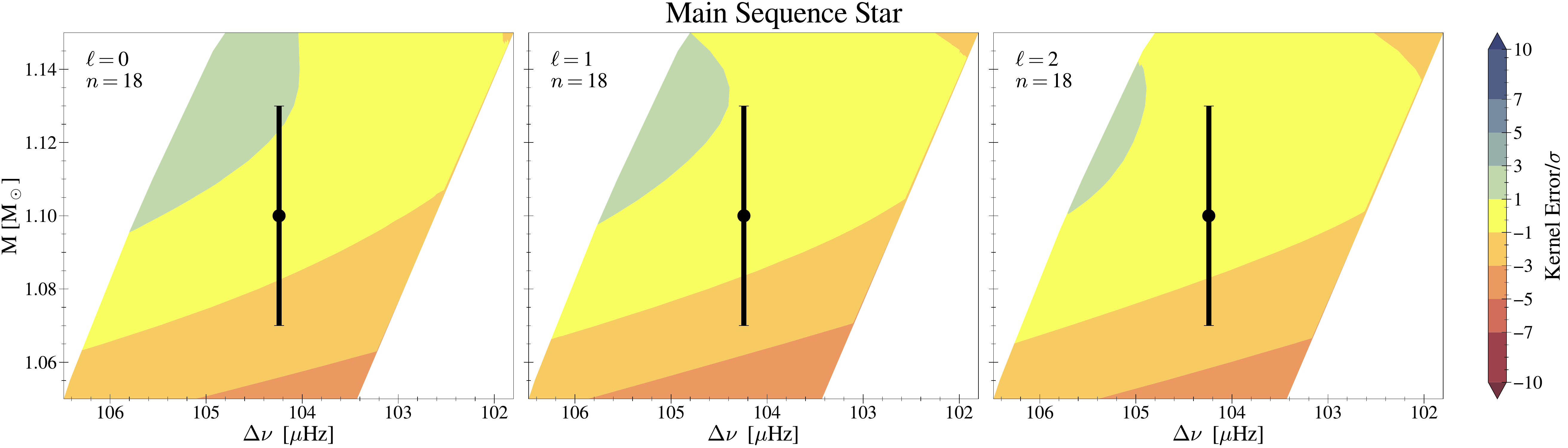}
    \caption{Contour plots of the kernel errors, scaled by a representative uncertainty ($\sigma = 10^{-4}$), of a main sequence star for the modes with frequencies nearest $\nu_{\rm{max}}$. The $x$-axis, which shows $\Delta \nu$ as a proxy for stellar age, is reversed so that a model of a given mass evolves horizontally from left to right through the plot.  The point in the center of the plot shows the location of the reference model used and the error bars represent the typical uncertainties in mass and $\Delta \nu$, taken from \citet{2017ApJ...835..172L}. The uncertainty of $\Delta \nu$ is smaller than the point size. }
    \label{fig:MS_KE}
\end{figure}

In order to understand how the kernel errors vary across our grid, we take the model at the center of our grid ($M=1.1~\rm{M}_\odot$, $\nu_{\rm{max}} = 2205~\rm{\mu Hz}$) as our reference model. For the 13 radial orders with frequencies closest to $\nu_{\rm{max}}$, we calculate the kernel error between our reference model and every model in the grid using Equation~\ref{equ:kern_err}. Figure~\ref{fig:MS_KE} shows our results for the central radial order as a function of the large frequency separation  $\Delta \nu$, which is a proxy for the stellar age. In this figure, the color shows the kernel error divided by $\sigma = 10^{-4}$, which corresponds to the typical uncertainty of the relative frequency differences \citep{2016MNRAS.456.2183D, 2017ApJ...835..172L}. Here we see that a large part of the parameter space has kernel errors that are below the observational uncertainties. The largest differences occur with models that are lower in mass and lower in $\Delta \nu$ (and therefore older) than our reference model. While Figure~\ref{fig:MS_KE} shows only one radial order, we obtain similar results for all 13 radial orders we explored. This is due to the pure acoustic nature of all the modes and the fact that the eigenfunctions of acoustic modes remain fairly constant across evolution and different masses. 

\subsection{OLA Inversions} 
Individual mode kernels are necessary for structure inversions, however, the final result is obtained through a linear combination of frequency differences. Thus, it is important to understand how these kernel errors propagate through an inversion. For this, we perform a set of model-model inversions using four different test models. Two test models are taken from the grid used to calculate the kernel errors, the model with the largest average kernel error (5.4$\sigma$) and one with a small average kernel error (1.3$\sigma$). For the third test model, we keep the same mass and $\Delta \nu$ but vary the initial composition such that the surface value of [Fe/H] is 0.1 dex higher, which is the typical uncertainty of spectroscopic [Fe/H] determinations of this type of star \citep[e.g.,][]{Mathur_2017, Furlan_2018}. The final test model was constructed with the same metallicity as the reference model but with the individual metal fractions scaled to the \citet{2009ARA&A..47..481A} abundance measurements instead of the \citet{1998SSRv...85..161G} scaling used in the reference model. This model also uses high-temperature opacity tables from OP \citep{2005MNRAS.362L...1S} instead of OPAL \citep{Iglesias1993, Iglesias1996}. The properties of all models used in our test inversions are given in Table~\ref{tab:ms_mod_prop}. 

\begin{deluxetable}{lccccc}[p]
\tablecaption{Properties of models used for main-sequence test inversions} 
\label{tab:ms_mod_prop} 
\tablehead{\colhead{Model} & \colhead{$M$/M$_\odot$} & \colhead{$\Delta \nu$/$\mu$Hz} & \colhead{[Fe/H]} & \colhead{$X_c$} & \colhead{Change}} 
\startdata
Reference Model & 1.1 & 104.42 & 0.1&  0.107 &\nodata \\
High Kernel Error & 1.05 & 103.5 & 0.1 & 0.015& Different $M$ and $\Delta \nu$ \\ 
Low Kernel Error & 1.093 & 103.23 & 0.1 & 0.065 & Different $M$ and $\Delta \nu$ \\
Different Z & 1.1 & 105.02 & 0.2 & 0.041 &  Composition \\ 
A09 Composition & 1.1 & 104.36 & 0.1 &0.079&  Metallicity fraction and opacity table source 
\enddata
\end{deluxetable} 

\begin{figure}[p]
    \epsscale{1.1}
    \plotone{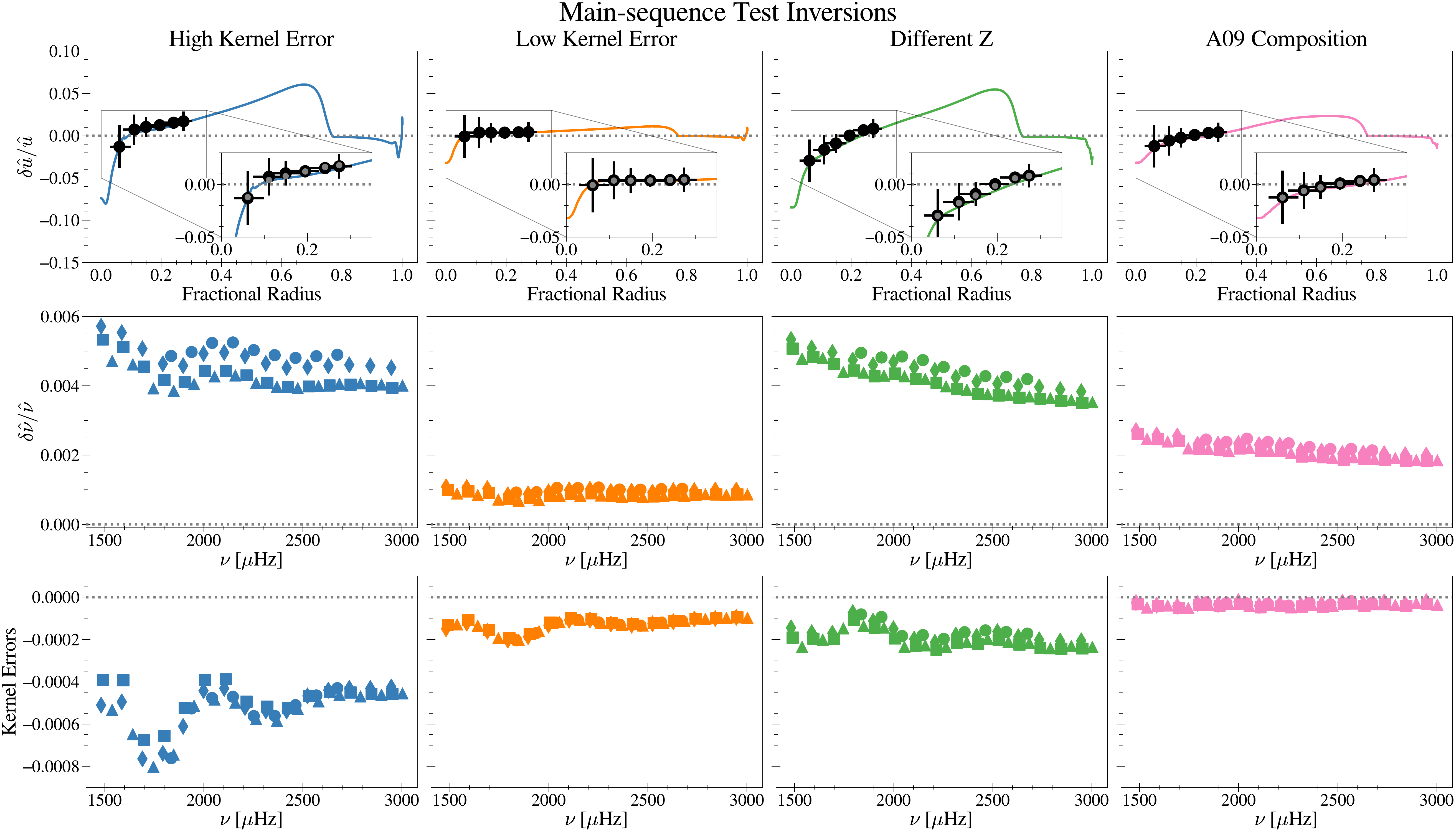}
    \caption{Results of several model-model test inversions. In all cases the same reference model and inversion parameters were used, and only the test model was varied. The properties of each test model are given in Table~\ref{tab:ms_mod_prop}. The first row shows the inversion results, the colored line represents the true difference in $\hat{u}$ and the black points are the inversion results. The vertical error bars show the uncertainty calculated using Equation~\ref{equ:inv_uncertain}. The horizontal error bars represent the FWHM of the averaging kernel. \rev{In some cases, particularly the innermost points, the peak in the averaging kernel is asymmetric, resulting in asymmetric horizontal error bars.}  The insets zoom into the region where inversions are sensitive. The insets also show the localized average difference as defined in Equation~\ref{equ:lad} as gray points. \rev{We do not show the $\delta Y$ profile as this term is too small to contribute to the frequency differences}. The second row shows the relative dimensionless frequency differences between the two models. The last row shows the kernel error of each mode. In the bottom two rows the symbol used indicates the spherical degree of the mode with squares (triangles, diamonds, circles) representing $\ell=$ 0 (1,2,3). In all cases the inversions yield accurate inferences of the internal stellar structure, even in the case of high kernel error.}
    \label{fig:MS_INV}
\end{figure}

For our inversions, we adopt the observed mode set and uncertainties of 16~Cyg~A from \citet{2017A&A...604A..42R}. We use this mode set because 16~Cyg~A has a similar mass and $\nu_{\rm{max}}$ as our reference model. This mode set also represents a best-case scenario due to the large number of modes observed, including $\ell=3$ modes, and the low uncertainties of the measured frequencies. In Figure~\ref{fig:MS_INV} we show the results of these test inversions as well as the frequency differences and kernel errors between our reference and test models. 

\rev{Before discussing the inversion results, it is interesting to discuss the behavior of the kernel errors between different test models. Since Equation~\ref{equ:kern_err} is correct to first-order, the leading order in the kernel errors shown in Figure~\ref{fig:MS_INV} may be the second-order term due to the structure differences. We can roughly estimate this by comparing the frequency and structure differences between the ``Low Kernel Error'' and ``Different'' cases. Both the relative frequency differences and the relative structure differences of the ``Low Kernel Error'' case are roughly one fifth of the ``Different Z'' case. One might expect then, that the kernel errors of the ``Low Kernel Error'' case should be roughly one twenty-fifth ($= (1/5)^2$) of the kernel errors of the ``Different Z'' case. However, Figure~\ref{fig:MS_INV} shows that the kernel errors of the two cases are almost the same. This is because the previous analysis only accounts for the second-order effect of the structure differences. There are, however, other potential sources of second-order effects. These include changes due to differences in the eigenfunctions of the modes, as well as,  or second-order effects in the change of variables used to construct the ($\hat{u},Y$) kernels. That is, the expression used to derive the kernels $\delta \hat{u}/\hat{u} = \delta \hat{P}/\hat{P} - \delta \hat{\rho}/\hat{\rho}$, where $\hat{P}$ is the dimensionless pressure and $\hat{\rho}$ is the dimensionless density, is only true to first-order.}

\rev{Returning to our test inversions, we find in all four cases} that our inversions perform well, with the inversion results closely matching both the true structure differences and the localized average difference. This is in line with previous work on solar inversions \citep{2000ApJ...529.1084B}. Somewhat surprisingly, even the inversions using a test model with very high average kernel errors still recover the correct differences. We attribute this to the fact that the kernel errors are roughly constant across different modes, and thus the linear combination formed during the inversion procedure acts to suppress the total error. The result is that kernel errors propagated through the inversion procedure are much smaller than the propagated mode uncertainties. We discuss this in more detail in Appendix~\ref{app:kern_err_prop}. 

We also performed the same analysis using a grid of models with convective cores and present these results in Appendix~\ref{app:cov_core_ms}. From these tests, we conclude that the range of linearity for structure inversions on the main sequence is quite broad. This is due to two factors. First, the stability of p-mode kernels across varying masses, ages, and compositions means that there is a large range of the parameter space with intrinsically low kernel errors. Secondly, the kernel errors across different modes behave similarly, meaning that the overall effect of these errors is suppressed by the structure inversion procedure.

\section{Subgiant Stars} 
\label{sect:SGB} 
Having \rev{examined} the reliability of structure inversions of main-sequence solar-like oscillators, we now turn to stars on the subgiant branch. For this we keep the same mass values \rev{and composition} as our main-sequence grid, however now we examine more evolved models with $\nu_{\rm{max}}$ between range 1282 and 1150~$\mu$Hz. Again we use the model at the center of our grid (1.1~M$_\odot$, $\nu_{\rm{max}} = 1216~\mu$Hz) as our reference model.  
\subsection{Kernel Errors} \label{sect:SGB_mode_discuss}
\begin{figure} 
    \plotone{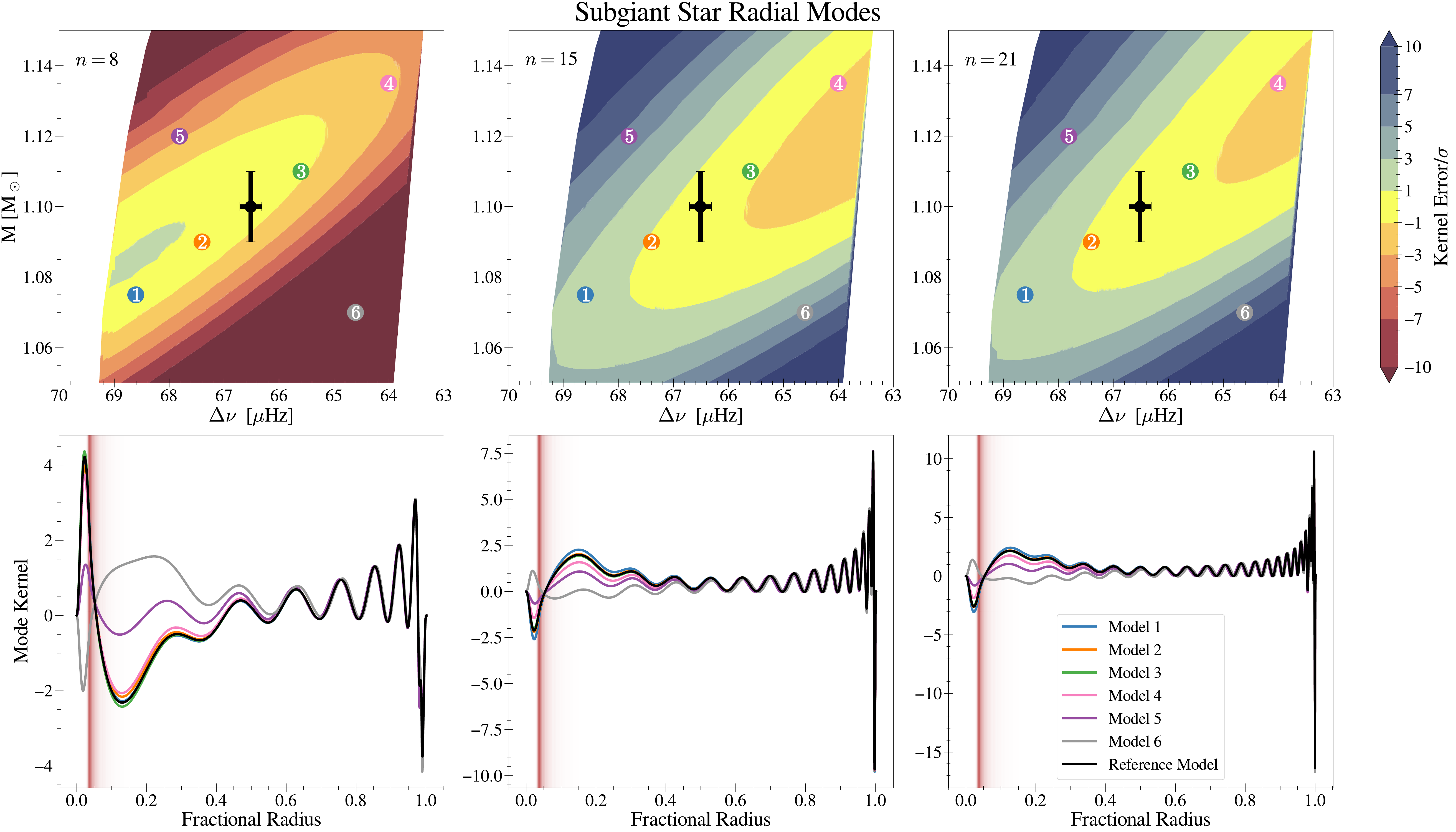}
    \caption{Contour plots of kernel errors for several radial modes of our subgiant grid. The reference model is indicated with a black point and the error bars correspond to the uncertainty of mass and $\Delta \nu$ of $\mu$Her given in \citet{2017ApJ...836..142G}. We also show, in the bottom row, the mode kernels of our reference model and several other models, indicated in the top row with colored points and numbers. The red vertical shading indicates the strength of nuclear burning in the hydrogen burning shell of the reference model. The kernels of model 6 (shown in gray) have opposite signs in the near-core region because this model has already passed through the singularity found by \citet{2021ApJ...915..100B}. } 
    \label{fig:SGB_rad_KE}
\end{figure}

We repeat the same procedure to calculate the kernels errors of $\ell=0,1,2,3$ modes for 14 radial orders, centered around $\nu_{\rm{max, ref}}$. In Figure~\ref{fig:SGB_rad_KE}, we show the results of several radial modes across different radial orders. For these modes, the regime where the kernel errors are less than our uncertainty is smaller than in the case of the main-sequence models. However, it is still large enough that fitting the mass and $\Delta \nu$ within 1$\sigma$ returns a model with low kernel errors. This is unsurprising, as the radial modes are purely acoustic and so benefit from the same stability in the eigenfunctions as was seen in our main-sequence models. We do however find slightly larger differences at the edges of our grid, particularly in the lower order modes. To better understand where these errors arise from, we also plot the mode kernels of several different models from our grid in the lower panels of Figure~\ref{fig:SGB_rad_KE}. 

Kernel errors arise when the mode kernel of the reference model no longer approximates the mode kernel of the test model. Thus we would expect, and indeed find, that the test models with the highest kernel errors (models 5 and 6) show the greatest difference between the mode kernels of the test and reference models. There is some variation in the kernel error contours across different radial orders. Most notably, the lower order mode shows negative kernel errors on the outer portions of the grid in contrast to the two higher order modes. This is explained by looking at the mode kernels. The mode kernel of the lower order mode is positive below the hydrogen burning shell and negative above it. The higher order modes show the opposite behavior. Across all radial orders, models which are within the observational uncertainties of mass and $\Delta \nu$ show small kernel errors for the radial modes.

\begin{figure} 
    \plotone{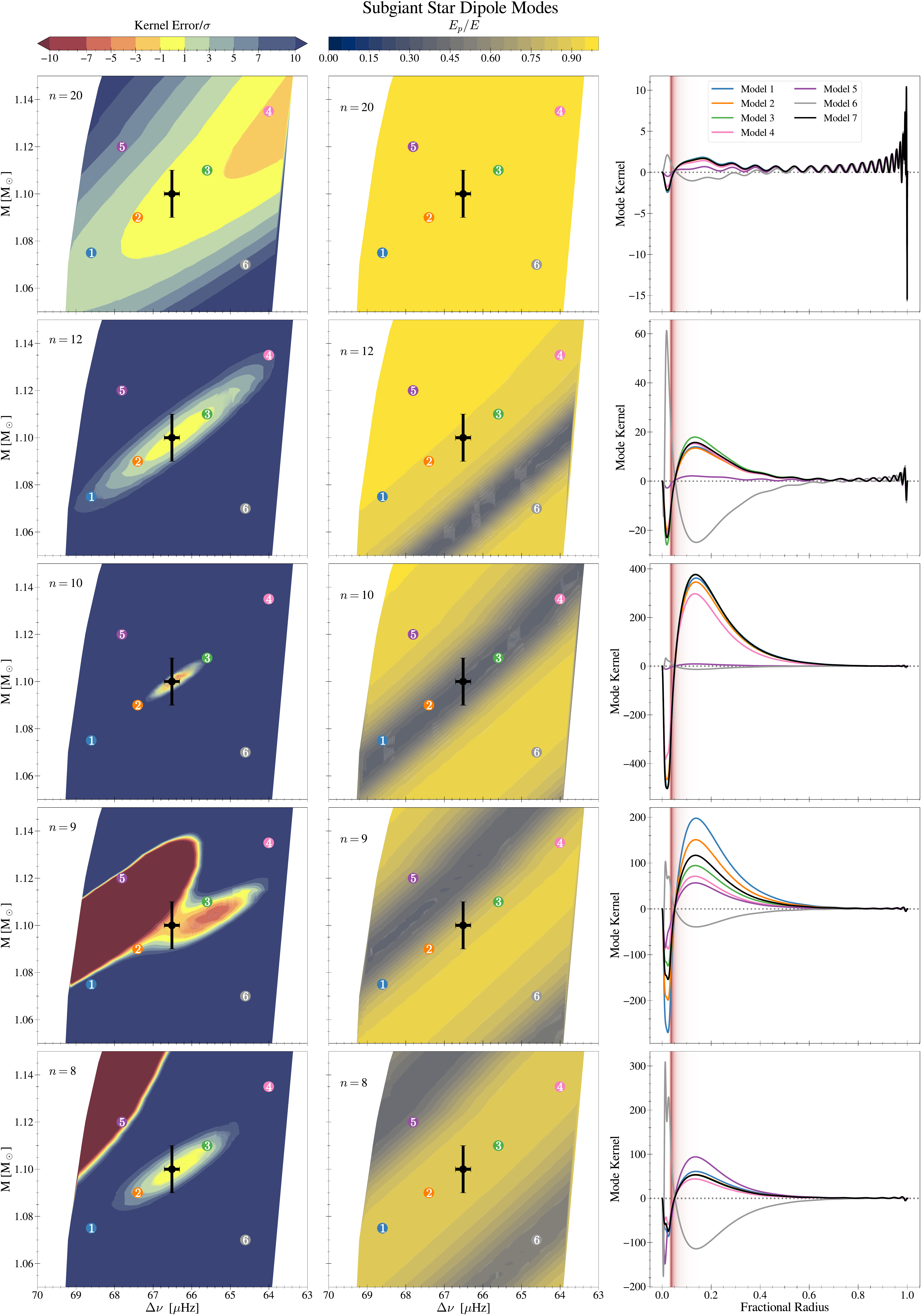}
    \caption{Left column: same as the top row of Figure~\ref{fig:SGB_rad_KE} but for several dipole modes of the subgiant star. We indicate the radial order of each mode. This is the value reported in GYRE as \texttt{n\textunderscore pg} and is defined using the Eckart-Scuflaire-Osaki-Takata scheme \citep{2006PASJ...58..893T}. Center column: mode inertia of the p-mode cavity, which indicates the mode character of the mixed mode. Right column: mode kernels of our reference model and several other models indicated in the left column with colored points and numbers. The red vertical shading indicates the strength of nuclear burning in the hydrogen burning shell of the reference model.}
    \label{fig:SGB_dip_KE}
\end{figure}

The situation changes dramatically for the dipole modes, shown in Figure~\ref{fig:SGB_dip_KE}. When discussing the behavior of mixed modes, it is useful to know the mode character, or how p- or g-dominated the mixed mode is. To visualize this, Figure~\ref{fig:SGB_dip_KE} also shows the acoustic mode inertia. This quantity is defined as
\begin{equation}
    \label{equ:p-inerita} 
    \frac{E_p}{E} = \frac{\int_{\textrm{p-cavity}} \left[\xi^2 + l(l+1) \eta^2\right] \rho r^2 \textrm{d}r}{\int_{0}^R \left[\xi^2 + l(l+1) \eta^2\right] \rho r^2 \, \textrm{d}r}, 
\end{equation}
where \(\xi\) and \(\eta\) are the radial and horizontal components of the eigenfunction, respectively, and the integral in the numerator is evaluated only over the acoustic cavity\footnote{This is the region of the star where $S^2_\ell < \omega^2$ and $N^2 < \omega^2$.   Here $S_\ell = \ell(\ell+1)c^2/r^2$ is the Lamb frequency, $\omega = 2 \pi \nu$ is the angular frequency, \rev{and $N^2 = g \left(\frac{1}{\Gamma_1} \derfrac{\, \ln P}{r} - \derfrac{\, \ln \rho}{r}\right)$ is the Brunt-V\"{a}is\"{a}l\"{a} frequency.}}. A value of $E_p/E$ close to 1 corresponds to a mixed mode that is strongly p-dominated and, conversely, a value of $E_p/E$ close to 0 corresponds to a mode that is strongly g-dominated. We also show, in Figure~\ref{fig:SGB_dip_KE}, the mode kernels of several models within the grid. 

For these dipole modes, we see several different ways that the kernel errors change over the parameter space. The easiest of these to explain are the highest order modes, which behave in a manner that is similar to the radial modes. This is expected as these modes have not yet experienced an avoided crossing at any point in the parameter space and so fully retain their acoustic character. 

For the mixed dipole modes the most important feature of the kernels is the amplitude of the two peaks in the near-core region. As a mode evolves through an avoided crossing the amplitudes of these two peaks increases as the mode becomes more g-dominated, and hence more sensitive to the core. This is why the regions of linearity lie along lines of constant acoustic inertia. At the beginning and end of the avoided crossing the mode character changes slower than in the middle, which results in narrower regions of linearity for modes that are strongly g-dominated in the reference model, i.e., $n=10$. The regions where the kernel error is negative seem to be a combination of the kernel amplitude and the structure differences with a small region of linearity where the kernel amplitude matches that of the reference model. The kernel errors of higher degree modes ($\ell = 2,3$) show similar behavior to the dipole modes, we discuss them in more detail in Appendix~\ref{app:sgb_details}. We also discuss in this appendix the singularity in the equations used to obtain the $(\hat{u},Y)$ kernels noted in  \citet[][Appendix A]{2021ApJ...915..100B}.


\subsection{OLA Inversions} 
\begin{figure}[p]
    \epsscale{1.1}
    \plotone{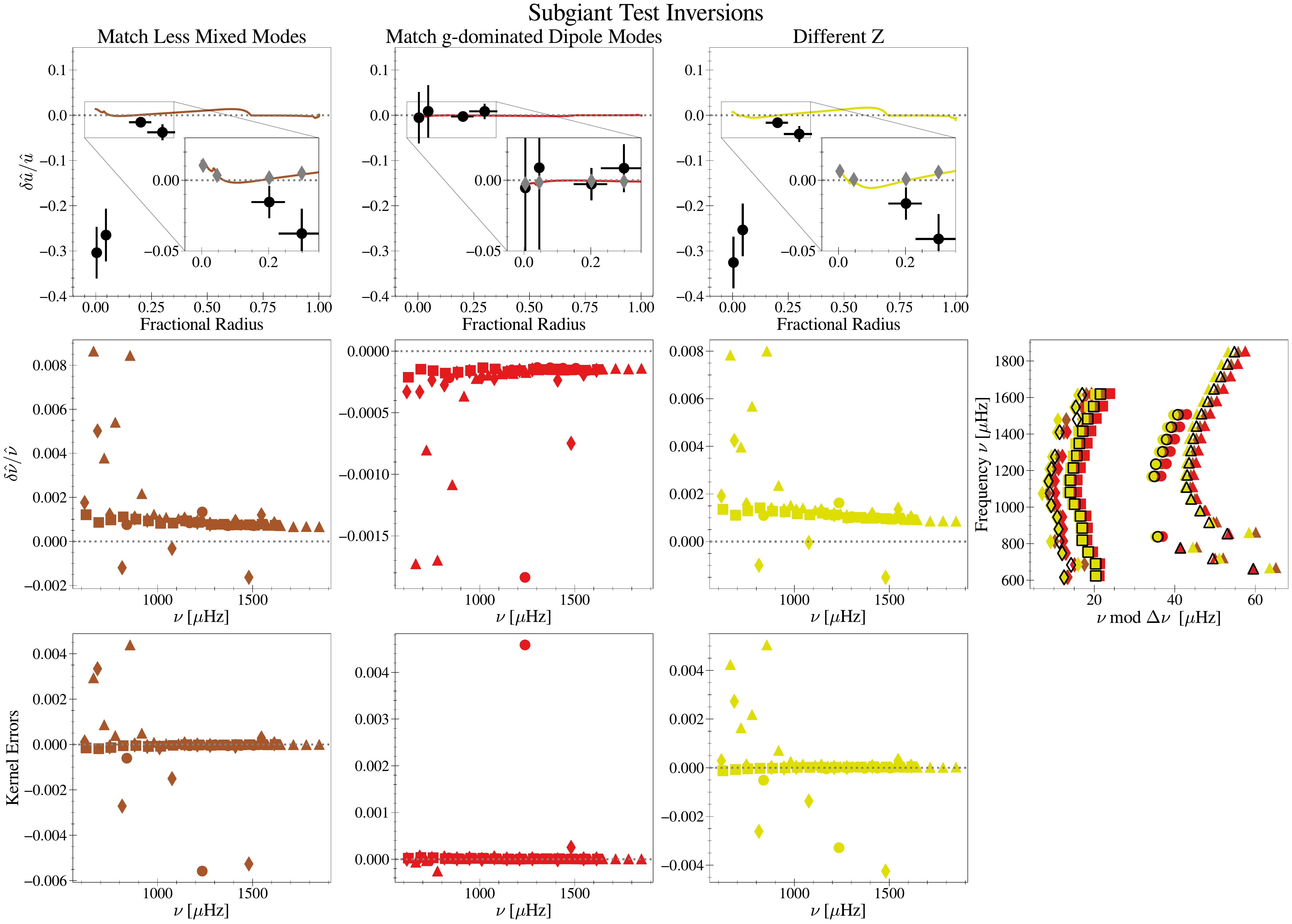} 
    \caption{Results of several model-model test inversions for the subgiant case. In all cases the same reference model and inversion parameters were used, and only the test model was varied. The properties of each test model are given in Table~\ref{tab:sgb_mod_prop}. The first row shows the inversion results, the colored line represents the true difference in $\hat{u}$ and the black points are the inversion results. The insets zoom into the region where inversions are sensitive and also show the localized average difference as gray diamonds. \rev{The horizontal uncertainties of the inversion results represent the width of the averaging kernel. For the innermost two target radii they are smaller than the size of the points.} The second row shows the relative dimensionless frequency differences between the two models. The last row shows the kernel error of each mode. The symbols indicate the spherical degree of the mode in the last two rows as in Figure~\ref{fig:MS_INV}. In the fourth column we also show the frequency \'echelle diagram of all models. The reference model is plotted in open black points and the test models are shown with fillled symbols according to the color used in the other columns.} 
    \label{fig:SGB_INV}
\end{figure}

Clearly the magnitude and morphology of kernel errors on the subgiant branch differ significantly from that of the main-sequence cases. We now seek to explore how these errors propagate through a set of inversions. Again, we select a set of test models, whose parameters are given in Table~\ref{tab:sgb_mod_prop}. We show the frequency \'echelle diagram of all models and the results of our test inversions in Figure~\ref{fig:SGB_INV}.  For these test models, we searched for cases where the agreement in the frequencies between our test and reference models are qualitatively similar to other best-fit subgiant models found in the literature \citep{2021A&A...647A.187N, 2021ApJ...915..100B, 2024ApJ...965..171L}. Note that the models selected as test models are not those used to show the variation of mode kernels in Figures~\ref{fig:SGB_rad_KE}~and~\ref{fig:SGB_dip_KE}. 

We take two of our three test models from the grid used to calculate kernel errors. The first was chosen to resemble the quality of the fit shown in \citet{2021A&A...647A.187N} and \citet{2024ApJ...965..171L}. This test model closely matches the radial mode frequencies, as well as many of the quadrupole and octopole modes, but has quite large differences in the mixed dipole modes. The second test model was chosen to resemble the quality of fit shown in \citet{2021ApJ...915..100B}, where the frequencies of the most mixed dipole modes are matched much better than the less mixed dipole modes. The third test model was constructed with the same mass as the reference model but a different composition. No other physics was changed and the $\Delta \nu$ value is similar to the reference model. In all cases, our test models are very close in mass (\(\pm 0.01\)M\(_\odot\)), well within the typical uncertainties of subgiant mass determinations \rev{\citep[e.g.,][]{2020MNRAS.495.3431L,2021A&A...647A.187N, 2024ApJ...965..171L}}. 

For these inversions, we adopt the observed mode set and uncertainties of $\mu$Her obtained using a recent release of SONG data (Kjelsen et al., in prep). The long observation baseline of the SONG project \citep{2007CoAst.150..300G} results in a very large number of modes identified with very small uncertainties. Thus, as with the mode set used in our main-sequence tests this represents the best-case scenario for structure inversions given current observations. 

In one case (Match g-Dominated Dipole Modes), we are able to recover the known differences within the uncertainties of the inversion. However this represents an extremely optimistic quality of fit between reference model and test model. \rev{This results in only one mode showing a high kernel error. This alone could introduce significant error into the inversion results, if the inversion coefficient of this mode was high. However, in this case it appears that the inversion coefficient of this mode is accidentally small compared to the other modes. Thus the systematics introduced by this high kernel error mode are small.}

In the remaining two cases, our test inversions are unable to recover the true differences between models. This is not due to the quality of the averaging kernels as our localized averaged differences are in agreement with the known differences. \rev{The fact that the results are similar in these two cases is unsurprising as all of the ingredients in the structure inversion (structure differences, frequency differences, inversion coefficients, and kernel errors) are similar. The large difference between the inversion results and the localized average differences result from both the high kernel errors and the importance of these modes to the construction of the averaging kernels. That is, these modes have high kernel errors and high inversion coefficients. We also do not see the suppression of the kernel errors found in the main-sequence case. This is due to the high variation in the kernel errors between different modes,} see Appendix~\ref{app:kern_err_prop} for more details.

\begin{deluxetable}{lccccc}[p]
\tablecaption{Properties of models used for subgiant test inversions} 
\label{tab:sgb_mod_prop} 
\tablehead{\colhead{Model} & \colhead{$M$/M$_\odot$} & \colhead{$\Delta \nu$/$\mu$Hz} & \colhead{[Fe/H]} & \colhead{$\rho_c$/g$\cdot$ cm$^{-3}$} & \colhead{Change}} 

\startdata
Reference Model & 1.1 & 66.45 & 0.1 & 1467& \nodata \\
Match Less Mixed Modes& 1.091 & 66.47 & 0.1 & 1580 & Different $M$ and $\Delta \nu$ \\ 
Match g-dominated Dipole Modes  & 1.101 & 66.55 & 0.1 & 1446 & Different $\Delta \nu$ \\
Different Z & 1.1 & 66.37 &  0.12 & 1571 & Composition \\ 
\enddata

\end{deluxetable} 

\section{Conclusions} 
\label{sect:conc} 
In this work we have explored the reliability of linear structure inversions for solar-like oscillators on the main sequence and subgiant branch. Using a grid of stellar models for each evolutionary stage, we calculated the kernel errors between a reference model and each model within our grid. To understand how these kernel errors propagate through an inversion we also performed inversions between our reference model and several test models. For our main sequence models, we found that the kernel errors were low across much of the parameters space. Our main sequence test inversions showed that even in cases with larger kernel errors, the inversions return reliable results. 

On the subgiant branch, where non-radial models can exhibit a mixed acoustic-buoyancy nature, the picture is quite different. Here we found that the part of the parameter space where kernel errors are lower than observational uncertainties is much smaller, particularly for g-dominated mixed modes. This causes our test inversions to return erroneous values, even in cases where the global parameters (mass, $\Delta \nu$, [Fe/H]) of the test model are well within observational uncertainties of the reference model. 

As the errors in the subgiant structure inversions are primarily due to large kernel errors in a few modes (see Appendix~\ref{app:kern_err_prop}), it may be possible to perform inversions with these modes removed from the mode set and recover the correct result. However, it remains unclear how to determine the modes that should be removed based only on observations. One could consider removing modes based on the frequency difference between model and observations, but this approach has several problems. First, the correlation between the frequency difference and the kernel error is, at best, only moderate. As an example, the Spearman correlation coefficient between the non-radial frequency differences and kernel errors of the models used in our inversion tests range from 0.03 to 0.5, which correspond to a negligible to moderate correlation. The second problem is that removing modes from the mode set also removes information from the inversion, and thereby reduces the quality of the averaging kernel which can be constructed. 

Future work on probing the interior structure of subgiant stars will likely require improvements to the forward modeling procedure and modifications to existing inversion techniques. 
\rev{On the modeling side, future efforts to find reference models for structure inversions should develop fitting methods that prioritize matching the character of mixed modes, similar to the method proposed by \citet{2025A&A...693A.274A} for finding reference models for rotational inversions of red giant stars.}

At the same time, it is clearly necessary to expand asteroseismic inversion techniques beyond those originally developed to study the Sun. This need has already been noted by \citet{2023A&A...675A..17V} who explored the application of structure inversions to stars which oscillate in pure g modes.  Several works \citep{2018Natur.554...73G, 2024A&A...686A.267F} have proposed non-linear inversions, which iteratively perturb a static model and solve the full oscillation equations, thereby including non-linear effects not accounted for in the linear variational approach explored in this work. Alternatively, the variational approach could be expanded to account for higher order terms.

\begin{acknowledgments}
\rev{We would like to thank the anonymous referee for their insightful comments and constructive feedback.} The research leading to the presented results has received funding from the ERC Consolidator Grant DipolarSound (grant agreement \#101000296). We thank J. Christensen-Dalsgaard for comments on the early stages of this work and for sharing a manuscript, written in collaboration with Mike Thompson, describing similar work. Additionally, we thank F. Ahlborn, V. Vanlaer, and J. Ong for productive discussions.

\end{acknowledgments} 


\appendix
\section{Propagation of Kernel Errors} 
\label{app:kern_err_prop} 
\begin{figure} 
    \plottwo{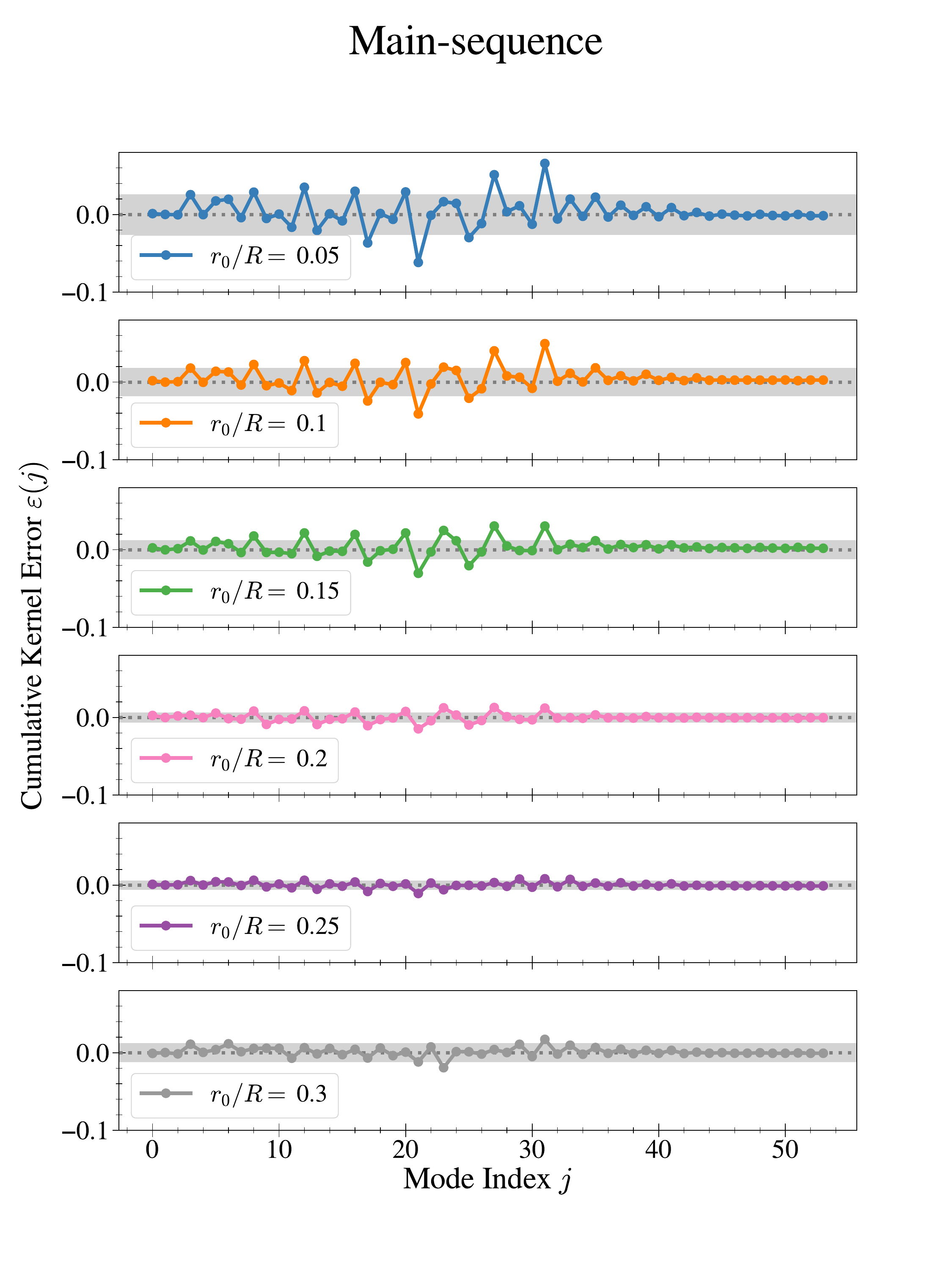}{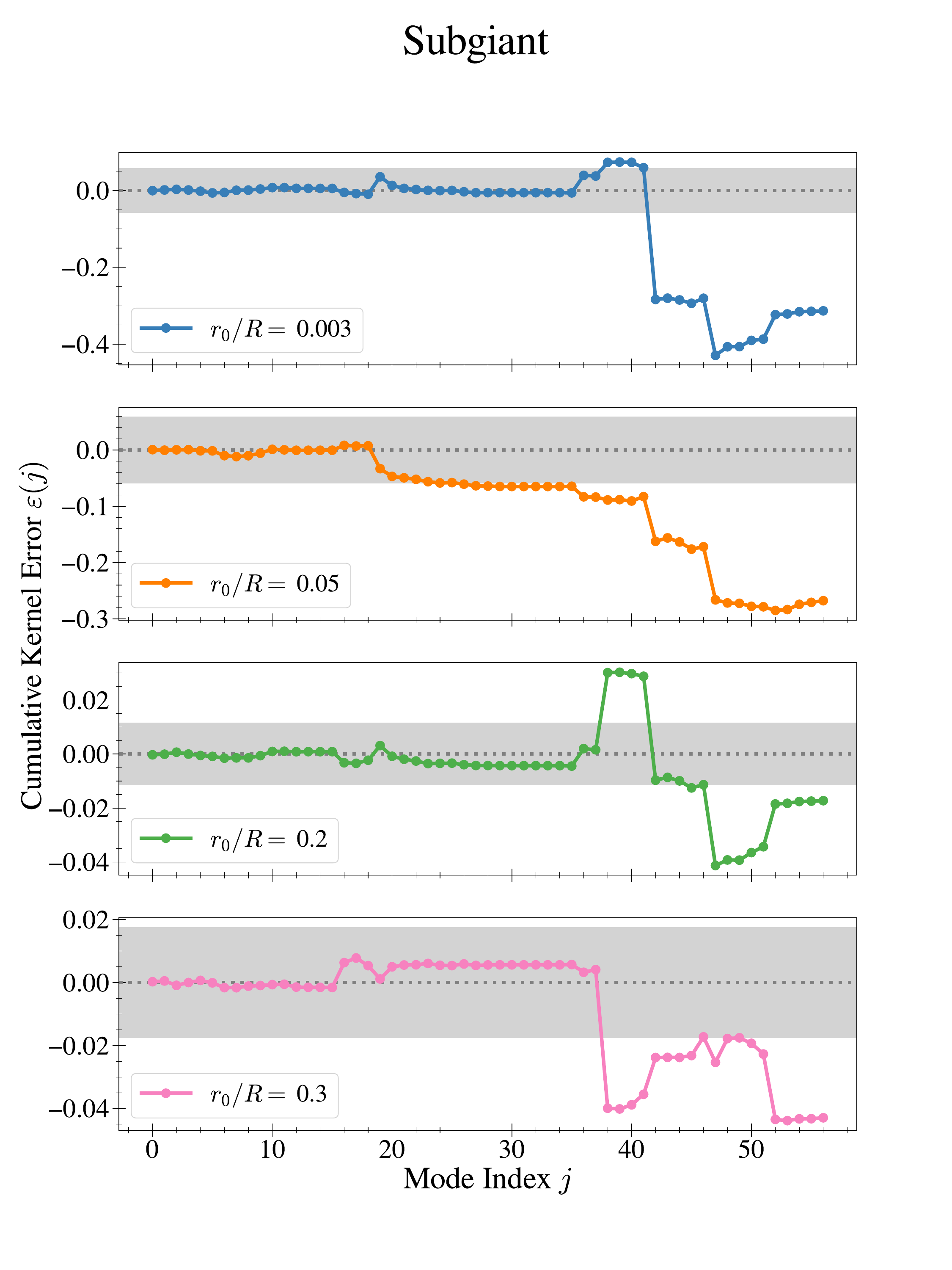} 
    \caption{Cumulative values of the kernel error, $\varepsilon(j)$, defined in Equation~\ref{equ:cumu_kern_err}, for our test inversions on the main sequence (left) and subgiant branch (right). The rightmost point in each panel represents $\varepsilon(N)$, the total error in the inversion due to the underlying kernel errors. The gray shaded region represents the uncertainty of the inversion result propagated from the uncertainty of the observed modes. The mode index values are discrete, however we connect the points to guide the eye.}
    \label{fig:inv_err}
\end{figure}

To understand why the high kernel errors cancel out in the test inversions of our main-sequence model but not our subgiant model, it is useful to see how the kernel error of each mode contributes to the overall result. To visualize this we define the quantity $\varepsilon$: 
\begin{equation}
    \label{equ:cumu_kern_err} 
   \varepsilon(j) = \sum_{i=0}^j c_i \, \textrm{KE}_i. 
\end{equation}
Here, the index variables $i,j$ correspond to a specific combination of $n$ and $\ell$, $\textrm{KE}_i$ is the kernel error of the $i$-th mode (calculated using Equation~\ref{equ:kern_err}), and $c_i$ is the corresponding inversion coefficient. $\varepsilon(j)$ represents the cumulative kernel error after $j$ modes have been added to the sum. Thus $\varepsilon(N)$, where $N$ is the total number of modes, represents the total kernel error term in the inversion. In Figure~\ref{fig:inv_err}, we plot $\varepsilon(j)$ for one set of test inversions in each evolutionary stage. For the main-sequence star the test model is the `High Kernel Error' model and for the subgiant star we use the `Match Less Mixed Modes' model. 

For the main-sequence test inversions across all target radii, the value of $\varepsilon(j)$  oscillates around zero and the final point, $\varepsilon(N)$, is smaller than the uncertainties of the inversion result. In the subgiant case however, there are a few modes which contribute much more to $\varepsilon$ than others resulting in values of $\varepsilon(N)$ that are greater than the uncertainties propagated from observations. 

\section{Main Sequence Stars with Convective Cores}
\label{app:cov_core_ms}
\begin{figure} 
    \epsscale{0.9}
    \plotone{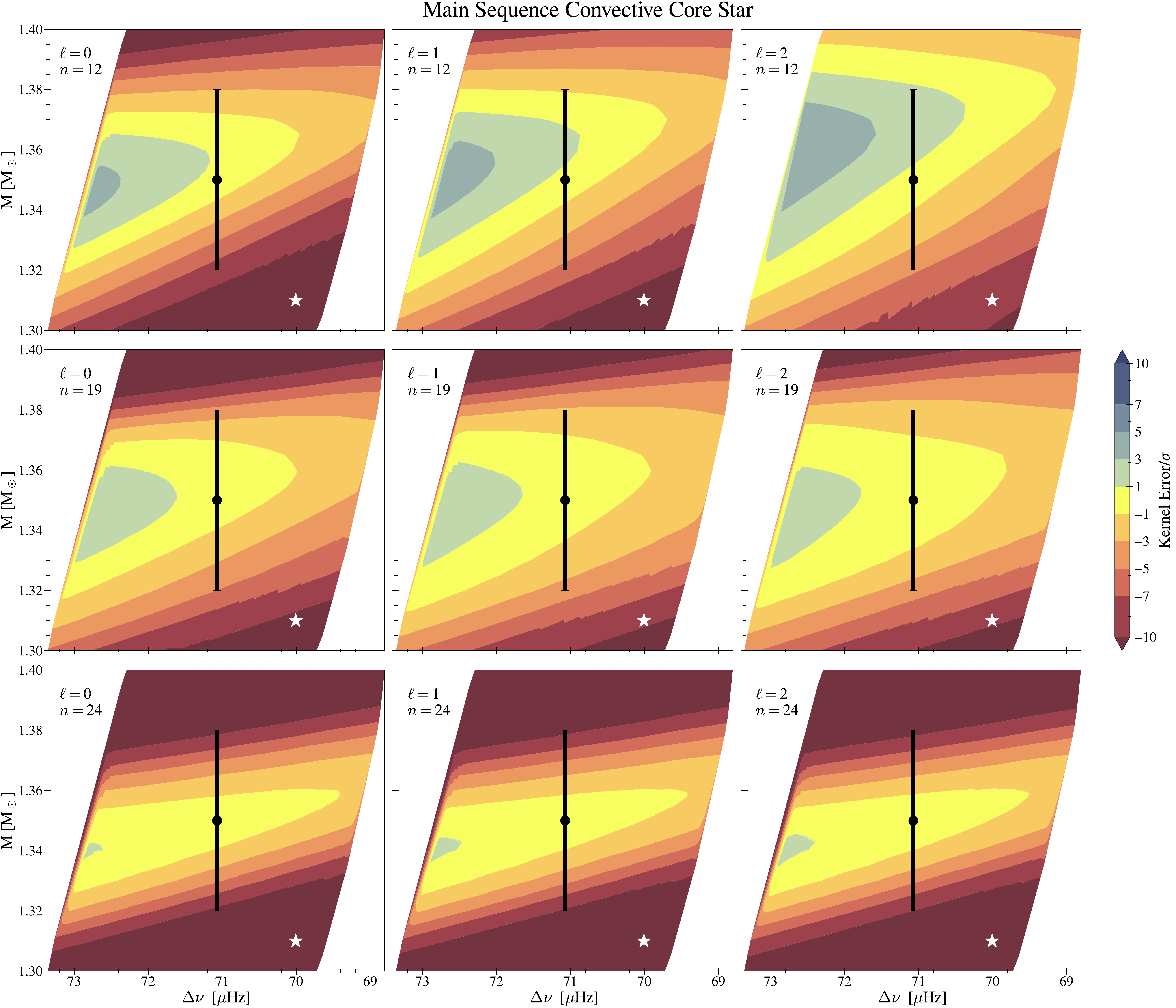}
    \caption{Kernel errors for a main-sequence star with a convective core. All colors and symbols have the same meaning as in Figure~\ref{fig:MS_KE}. In addition, the test model used for the test inversions presented in Figure~\ref{fig:MS_cc_INV} is indicated with a white star.}
    \label{fig:MS_cc_KE}
\end{figure}

\begin{figure} 
    \epsscale{0.4}
    \plotone{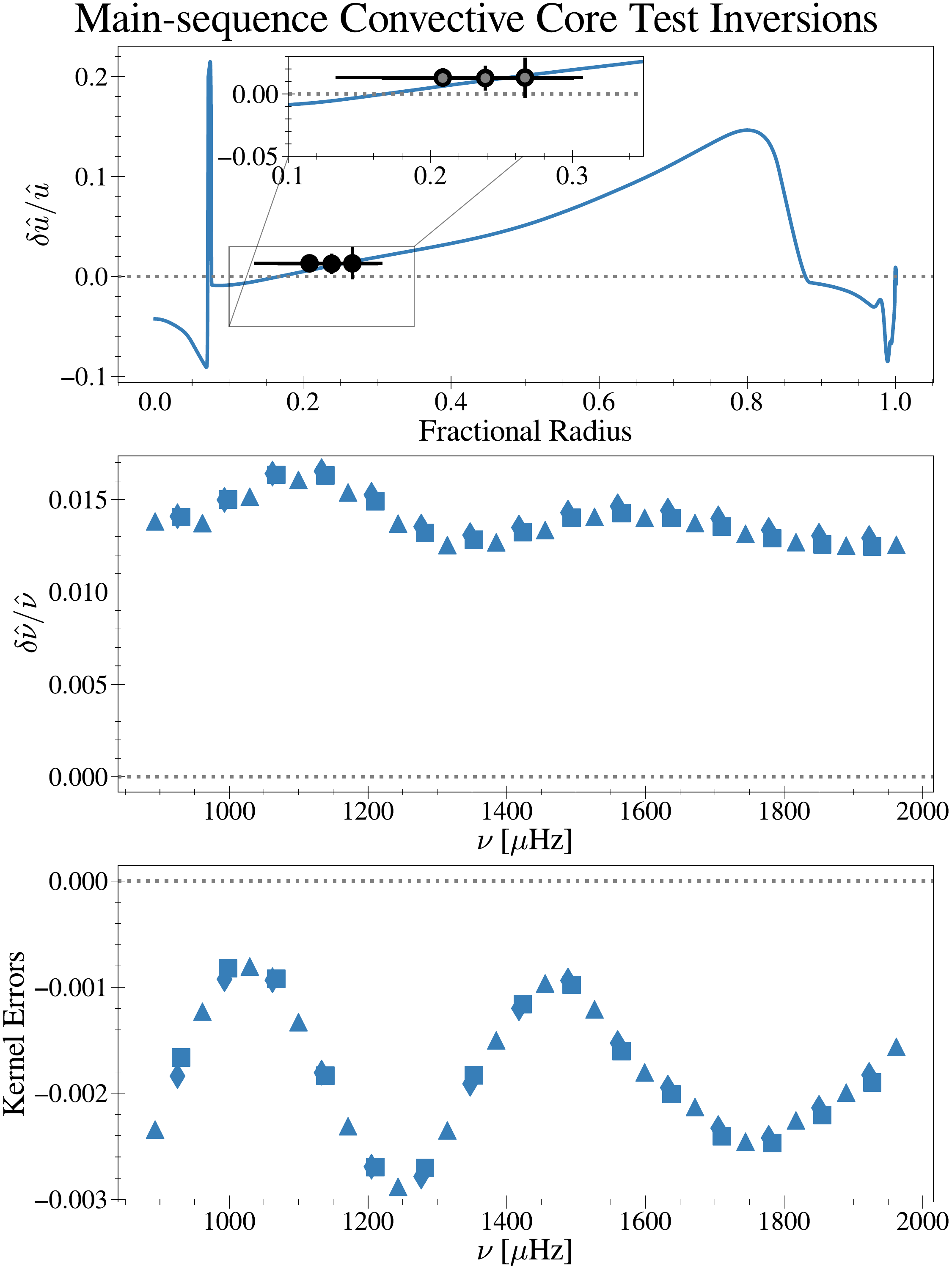}
    \caption{Results of a set of model-model test inversions for a model with a convective core. All symbols have the same meaning as in Figure~\ref{fig:MS_INV}.}
    \label{fig:MS_cc_INV}
\end{figure}

We repeat the calculation of the kernel errors for a 1.35M$_\odot$ main-sequence star with a convective core and show the results in Figure~\ref{fig:MS_cc_KE}. The region of low kernel errors is not quite as large as the radiative core case, discussed in Section~\ref{sect:MS}, particularity for the higher order models. We also select a test model with high average kernel errors and preform a set of model-model inversions and show the results in Figure~\ref{fig:MS_cc_INV}. For these inversions, we take the mode set of KIC~1435467 from \citet{2017ApJ...835..172L}. Despite the kernel errors being larger than the radiative core case, the inversion procedure still suppresses the errors and is able to infer the correct structure differences, given the quality of the averaging kernels. 

\section{Subgiant Higher Degree Modes and Singularity} \label{app:sgb_details} 

\subsection{Higher degree modes} \label{app:l23} 
\begin{figure} 
    \plotone{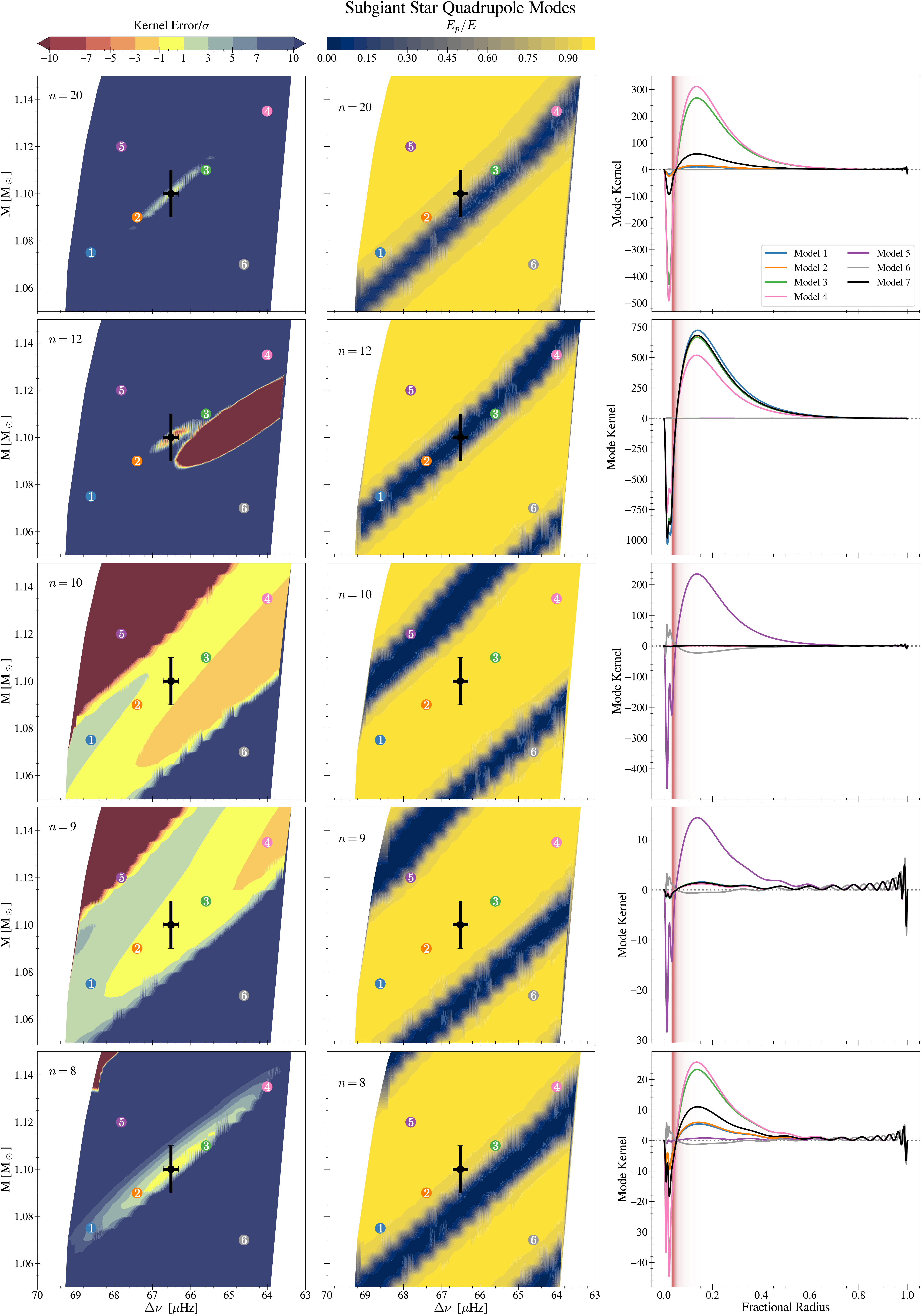}
    \caption{Kernel errors for the quadrupole modes of the subgiant stars. All colors and symbols have the same meaning as in Figure~\ref{fig:SGB_dip_KE}. }
    \label{fig:SGB_quad_KE}
\end{figure}

\begin{figure} 
    \plotone{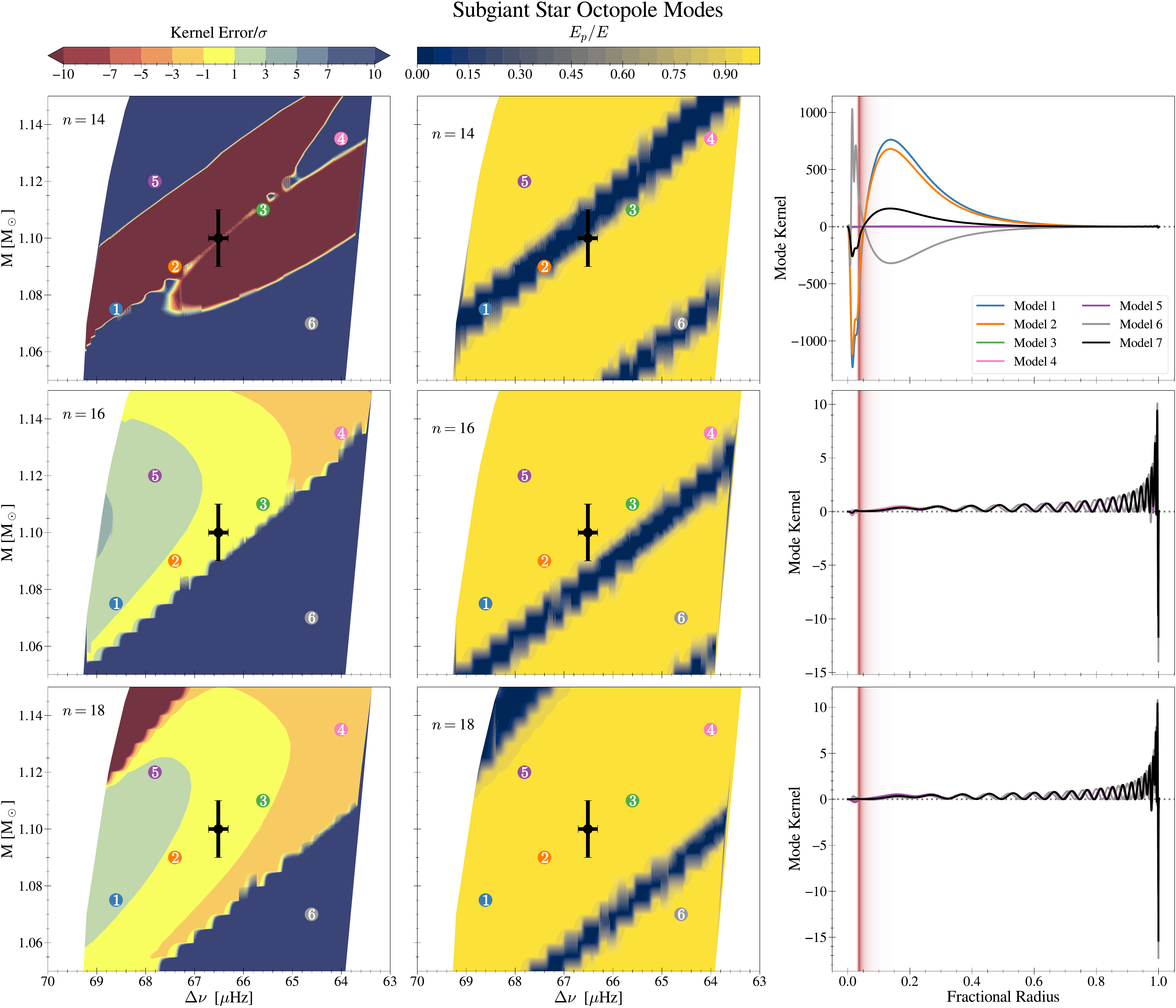}
    \caption{Kernel errors for the octopole modes of the subgiant stars. All colors and symbols have the same meaning as in Figure~\ref{fig:SGB_dip_KE}. }
    \label{fig:SGB_octo_KE}
\end{figure}

Here we show the kernel errors of our subgiant grid for modes of spherical degree $\ell=2$, Figure~\ref{fig:SGB_quad_KE}, and $\ell=3$, Figure~\ref{fig:SGB_octo_KE}. These modes fall into two categories based on the character of the mode in the reference model. For modes that are p-dominated in the reference model (high $E_p/E$) the kernel errors are low for test models that are in the same ridge of high $E_p/E$ and high for all other models. Modes that are g-dominated in the reference model show similar behavior to the g-dominated dipole modes discussed in Section~\ref{sect:SGB_mode_discuss}, although the region of linearity is smaller. This is due to the fact that higher-order modes evolve more quickly through avoided crossings. 

\subsection{Singularity} 
\citet{2021ApJ...915..100B} found a singularity in the equation used to obtain the $(\hat{u},Y)$ kernels. They derive that this singularity occurs when $\lambda =1$ where $\lambda$ is the eigenvalue of the homogeneous eigenvalue analogous to the differential equation used to obtain the $(\hat{u},Y)$ kernels. As a model evolves through this singularity the peaks of the mode kernels on either side of the hydrogen burning shell increase rapidly in amplitude and then change signs at the singularity, see Figure 4 of \citet{2021ApJ...915..100B}. This singularity affects all of the mode kernels of the model. However, it does not appear to affect the kernel errors as long as the reference model used is not passing through the singularity. 

\begin{figure}
    \plotone{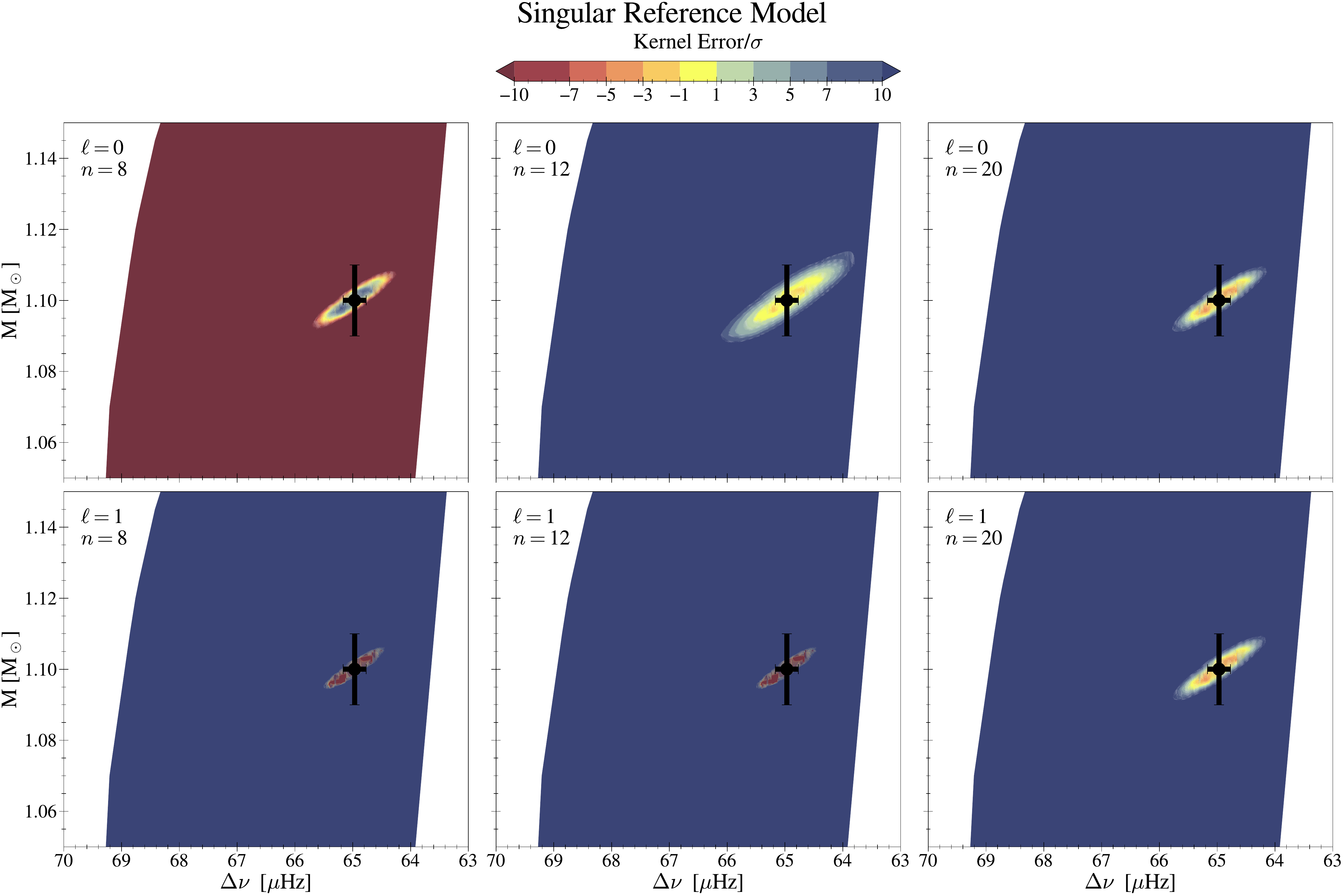}
    \caption{Kernel errors for several modes when the reference model used is passing through the singularity discussed in \citet{2021ApJ...915..100B}. The symbols have the same meaning as in Figure~\ref{fig:SGB_rad_KE}.}
    \label{fig:SGB_sing_KE}
\end{figure}

We also calculate the kernel errors using a reference model that is passing through this singularity $\lambda = 1.0001$. As seen in Figure~\ref{fig:SGB_sing_KE}, the region of linearity for this new reference model is extremely small even for the radial modes.  This is in agreement with the conclusion in \citet{2021ApJ...915..100B} that models with $|\lambda -1| \lessapprox 0.005$ should not be used for structure inversions.  

\clearpage
\bibliography{linearity}{}
\bibliographystyle{aasjournal}



\end{document}